\documentclass[12pt]{article}
\usepackage{amssymb,amsmath}
\usepackage{cite}
\newlength{\dinwidth}
\newlength{\dinmargin}
\newtheorem{theorem}{Theorem}[section]
\newtheorem{prop}[theorem]{Proposition}
\newtheorem{lemma}[theorem]{Lemma}
\newtheorem{cor}[theorem]{Corollary}

\newenvironment{proof}{\medskip \noindent 
            {\bf Proof.}}{ \hfill $\square$ \medskip}

\newcommand{\ie}{{\it i.e.\ }}

\def\msc{Modular Stability Condition}
\def\idty{{\leavevmode\hbox{\rm 1\kern -.3em I}}}

\def\wrnet{{\{\Rs(W)\}_{W \in\Ws}}}

\def\Bs{{\cal B}}
\def\Cs{{\cal C}}

\def\Hs{{\cal H}}

\def\Js{{\cal J}}

\def\Ls{{\cal L}}

\def\Os{{\cal O}}
\def\Ps{{\cal P}}

\def\Rs{{\cal R}}

\def\Ws{{\cal W}}

\def\Ys{{\cal Y}}
\def\Zs{{\cal Z}}

\def\Pid{{\Ps_+ ^{\uparrow}}}
\def\Lid{{\Ls_+ ^{\uparrow}}}

\def\idty{{\leavevmode\hbox{\rm 1\kern -.3em I}}}

\def\RR{{\mathbb R}}
\def\CC{{\mathbb C}}
\def\NN{{\mathbb N}}
\def\IN{{\mathbb N}}

\def\beq{\begin{equation}}
\def\eeq{\end{equation}}
\begin{document}
\title{Geometric Modular Action and Spontaneous Symmetry Breaking}
\author{{\Large Detlev Buchholz\,$^a$ \ and \  
Stephen J.\ Summers\,$^b$ }\\[5mm]
${}^a$ Institut f\"ur Theoretische Physik, 
Universit\"at G\"ottingen, \\ 37077 G\"ottingen, Germany  \\[2mm]
${}^b$ Department of Mathematics, 
University of Florida, \\ Gainesville FL 32611, USA}

\date{\small Dedicated to the memory of Siegfried Schlieder} 

\maketitle 

{\abstract \noindent We study spontaneous symmetry breaking for
field algebras on Minkowski space in the presence of a condition of
geometric modular action (CGMA) proposed earlier as a selection
criterion for vacuum states on general space--times. We show that any
internal symmetry group must commute with the representation of the
Poincar\'e group (whose existence is assured by the CGMA) and 
each translation-invariant vector is also Poincar\'e invariant.
The subspace of these vectors can be centrally decomposed into pure 
invariant states  and the CGMA holds in the 
resulting sectors. As positivity of the energy is not assumed,
similar results may be expected to hold for other space--times.}

\section{Introduction}

\setcounter{equation}{0}

   There are a number of physically relevant mechanisms which entail a
degeneracy of the vacuum state in quantum field theory. Primary among
these is the spontaneous symmetry breaking of an internal symmetry
group. Initiated by Borchers and by Reeh and Schlieder, systematic
study \cite{Bor1,Bor2,Ma1,Ma2,ReSch,Ar1,DS} in quantum field theories
satisfying the Wightman axioms \cite{StWi} or the standard axioms of
algebraic quantum field theory \cite{HaKa,Haag} has shown that the
presence of multiple vacua determines much of the global structure of
the theory. Common to these approaches is the assumption of the
positivity of the energy, with its concomittant analyticity
properties. In Minkowski space the spectrum condition is a natural and
physically meaningful assumption, but in other space--times it is
neither.
     
     It is therefore of interest to revisit both spontaneous symmetry
breaking and the structural consequences of degenerate vacua with the
standard axioms for Minkowski space theories replaced by a recently
proposed condition of geometric modular action (CGMA)
\cite{BS1,BDFS}. This condition is designed to characterize those
elements in the state space of a quantum system which admit an
interpretation as a ``vacuum''. It is expressed in terms of the
modular conjugations associated to the state and given family of
algebras indexed by suitable subregions (wedges) of the underlying
space--time and, in principle, can be applied to theories on any
space--time manifold.  For a motivation of this condition and
applications to theories in Minkowski, de Sitter, anti-de Sitter and a
class of Robertson--Walker space--times, we refer the interested
reader to \cite{BDFS,BFS,BMS1,BFS1,BMS2,AdS}. In this paper we shall
restrict our attention to four-dimensional Minkowski space, but the
arguments are applicable to other space--times, yielding similar
results.

     We shall consider an arbitrary group $G$ as the internal symmetry
group of a quantum field theory formulated in the algebraic context
\cite{Haag}. Hence, we shall assume there exists a net $\wrnet$ of von
Neumann algebras indexed by the set $\Ws$ of all wedges (Poincar\'e
transforms of the set 
$\{ x = (x_0,x_1,x_2,x_3) \in \RR^4 \mid x_1 > \vert x_0 \vert \}$) in
Minkowski space and acting upon a separable Hilbert
space $\Hs$, and that there exists a unitary representation $V$ of the
group $G$ such that
$$V(g) \Rs(W) V(g)^{-1} = \Rs(W) \quad , \quad g \in G \, , W \in \Ws \, .$$
We shall assume that there is a vacuum vector $\Omega_0 \in \Hs$
invariant under $V(G)$ but make no assumption about the invariance
properties of the other vacua. Indeed, one of the situations we are
interested in including in our analysis is the case where the 
various vacua are permuted among themselves by the action of $V(G)$.

     After specifying the working assumptions of this paper in Section
2, we shall show that in the presence of the CGMA, the internal
symmetries must commute with the representation of the Poincar\'e
group, whose existence is assured by the CGMA and which is constructed
using the modular conjugations.  In Section 3 we shall investigate the
global structure of the observable algebras and prove that any
translation-invariant vectors must also be Poincar\'e-invariant, in
contrast to what is known about vectors invariant under representations
of the translation group which satisfy the spectrum condition but do
not arise from modular objects \cite{Ar1,DS}. We then prove that under
the central decomposition of the global observable algebra all
relevant structures are preserved. Finally, in Section 4 we show that 
the CGMA and the modular stability condition introduced in \cite{BDFS}
manifest some remarkable rigidity properties.

\section{Modular action and internal symmetries}

\setcounter{equation}{0}

     Although the arguments presented here apply more generally, for
convenience we assume that the net $\wrnet$ is locally generated in
the sense defined in \cite{BS2} with a generating family $\Cs$ of
convex compact spacetime regions $\Os$. Roughly speaking, this means
that every algebra $\Rs(W)$ is generated by the family of all algebras
$\Rs(\Os)$ with $\Os \in \Cs$ and $\Os \subset W$. This subsumes such
familiar examples as nets generated by algebras associated with the
set of double cones.  Note that nets affiliated with quantum
field theories satisfying the Wightman axioms are locally
generated in this sense \cite{SWi}. For notational simplicity, we 
shall only consider bosonic theories here.

     We shall be assuming that the $V(G)$--invariant unit vector 
$\Omega_0 \in \Hs$ is cyclic and separating for $\Rs(W)$, for every 
$W \in \Ws$.\footnote{The fundamental insight that under physically motivated
conditions the vacuum vector is cyclic and separating for the quantum 
fields localized in wedge regions is due to Reeh and Schlieder
\cite{ReSch2}.} Thus, the Tomita--Takesaki modular theory will be
applicable, cf.\ \cite{BrRo,KaRi}. In the following $J_W$,
resp.\ $\Delta_W$, will denote the modular conjugation, resp.\ the
modular operator, associated to the pair $(\Rs(W),\Omega_0)$ by the
modular theory. Also, we shall use $\Js$ to represent the group
generated by the set $\{ J_W \mid W \in \Ws\}$.

     The following are included in the standing assumptions of this 
paper.

\begin{enumerate}
\item[(a)] $W \mapsto \Rs(W)$ is an order-preserving bijection.
\item[(b)] $\Omega_0$ is cyclic and separating for $\Rs(W)$, given 
any $W \in \Ws$.  
\item[(c)] For all $W_0,W \in \Ws$, 
$J_{W_0} \Rs(W) J_{W_0} = \Rs(\lambda_{W_0}W)$,
where $\lambda_{W_0} \in \Ps_+$ is the reflection through the edge
of the wedge $W_0$.
\end{enumerate}

\noindent In \cite{BDFS,BS2} the Condition of Geometric Modular Action
(CGMA), formulated solely in terms of the vector $\Omega_0$ and the net
$\wrnet$ without any {\it a priori} assumptions about the specific
form of the adjoint action of the modular conjugations on $\wrnet$ or
even the existence of an isometry group, was shown to entail
conditions (a)--(c). It has also been shown in \cite{BiWi} that (c)
must hold for any nets $\wrnet$ locally associated with
finite--component quantum fields satisfying the Wightman axioms. Note
that condition (c) implies that the adjoint action of any modular
conjugation $J_W$ leaves the set $\wrnet$ invariant.  As the
surjectivity of the map in (a) is automatic and the order preserving
property is just the operationally motivated condition of isotony,
only the significance of the injectivity assumption is not immediately
clear. It is shown in the Appendix that if the injectivity condition
is dropped, the remaining assumptions imply that the algebras $\Rs(W)$
are all abelian and independent of localization region $W$. Such a
situation is of no interest in quantum field theory. Hence, there is
no loss of physical generality to include in our standing assumptions
the requirement that in no subrepresentation of the net $\wrnet$ are
the wedge algebras abelian.

Condition (c) and modular theory imply 
$\Rs(W)^\prime = J_W \Rs(W) J_W = \Rs(W^\prime)$
for any wedge $W \in \Ws$, where $W^\prime \in \Ws$ denotes the causal
complement of $W$. Thus, the net fulfills wedge duality and hence 
\textit{a fortiori} locality. An immediate consequence 
of this fact is the following result about the type of the 
global algebra generated by  the wedge algebras. 
It is in perfect concord  with the idea 
that the CGMA characterizes elementary states. 

\begin{prop} \label{typeone}
Let $\wrnet, \, \Omega_0$ be a net and vector satisfying 
the standing assumptions, and let 
$\Rs = \bigvee_{W \in \Ws} \Rs(W)$. Then  
$\Rs^\prime \subset \Rs$ and $\Rs$ is of type I. 
\end{prop}
\begin{proof} 
Because of wedge duality, 
$\Rs^\prime = \bigwedge_{W \in \Ws} \Rs(W)^\prime
=  \bigwedge_{W \in \Ws} \Rs(W^\prime) \subset \Rs$.
Hence $\Rs^\prime$ coincides
with the center of $\Rs$, proving that $\Rs$ is of type  I. 
\end{proof}

     As shown in \cite{BDFS,BS2}, the standing assumptions also imply that
there exists a strongly continuous (anti)unitary representation $U$ of
the proper Poincar\'e group $\Ps_+$ on four-dimensional Minkowski
space, which is constructed in a canonical manner from products of the
modular conjugations $J_W \in \Js$ so that $U(\lambda_W) = J_W$, for
all $W \in \Ws$.  Indeed, one has $\Js = U(\Ps_+)$, so that $\Js$ is
closed in the strong-*-topology. One therefore has
$U(\lambda)\Omega_0 = \Omega_0$, for all $\lambda \in \Ps_+$. The
representation $U$ acts covariantly upon the net:
$$U(\lambda) \Rs(W) U(\lambda)^{-1} = \Rs(\lambda W) \, ,$$
for all $\lambda \in \Ps_+$, $W \in \Ws$. 

     Since the representation of the Poincar\'e group is constructed out
of modular involutions, a number of results which are difficult or not
possible to obtain in other settings follow easily in the presence of
the CGMA. Indeed, since $V(g)\Omega_0 = \Omega_0$ and 
$V(g) \Rs(W) V(g)^{-1} = \Rs(W)$, for all $g \in G$ and $W \in \Ws$, a 
basic result of modular theory entails that $V(g)$ commutes with all 
modular involutions $J_W$, $W \in \Ws$, cf.\ 
\cite[Corollary 2.5.32]{BrRo}. The commutation
of $V(G)$ with $U(\Ps_+)$ is therefore immediate.

\begin{theorem} 
If the standing assumptions are fulfilled,
$V(g)U(\lambda) = U(\lambda)V(g)$, for all $g \in G$ and $\lambda \in \Ps_+$.
\end{theorem}

      We note that the CGMA, and hence also the standing assumptions,
can be satisfied by examples in which the spectrum condition is
violated \cite{BDFS}. Landau and Wichmann showed that in the context
of a local net in an irreducible vacuum representation (with spectrum
condition) the internal symmetry group must commute with the
representation of the translation group \cite{LaWi}. With the further
assumptions that there is a mass gap in the theory and that for each
particle in the theory there exists a field with non-zero matrix
elements between the vacuum and the one-particle states, Landau proved
that the internal symmetry group must commute with the representation
of the Poincar\'e group \cite{La}. {}From another more technical
set of assumptions, Bisognano and Wichmann
\cite{BiWi2} were able to derive the same conclusion. Common to all
these earlier approaches is the assumption of the spectrum condition.

\section{Invariance and decomposition}

\setcounter{equation}{0}

Let $\Zs$ denote the center of the algebra 
$\Rs = \bigvee_{W \in \Ws} \Rs(W)$ and $\Zs(W)$ denote 
the center of $\Rs(W)$. Furthermore, let
$\Zs_{s}$ represent the set of all self-adjoint elements of $\Zs$. We
recall from Proposition \ref{typeone} that $\Zs = \Rs^\prime$ and
continue with some useful properties of these algebras.

\begin{prop} \label{centerinvariant}
Under the standing assumptions, $\Zs \subset U(\Pid)'$,
$\Zs_{s} \subset U(\Ps_+)'$ and $\Zs \subset \Zs(W)$,
for all $W \in \Ws$.
\end{prop}
{\bf Remark:} \ Since $\Zs^\prime = \Rs$, it follows from this result that
the unitaries $ U(\Pid)$ are elements of the global 
algebra $\Rs$, \ie the Poincar\'e transformations are weakly inner.
Again, this is in accord with the idea that the CGMA characterizes
elementary states.

\begin{proof}
As $\Zs = \Rs' \subset \Rs(W)' = \Rs(W')$ for any $W \in \Ws$,
one obtains $\Zs \subset \Zs(W)$ for all $W \in \Ws$.
But one knows from \cite[Lemma 3]{Ar2} that $J_W A J_W = A^*$, for 
all $A \in \Zs(W)$. Since every element of $U(\Pid)$ is a product 
of an even number of modular conjugations and every element of 
$U(\Ps_+)$ is the product of $J_W$ and an element of $U(\Pid)$ 
\cite{BDFS}, the remaining claims follow at once.
\end{proof}

     Let $E_0$ be the orthogonal projection onto the subspace 
of $\Hs$ consisting of $U(\RR^4)$--invariant vectors. Hence, we have 
$\Omega_0 \in E_0 \Hs$. It therefore follows from the preceding
result that 
$\overline{\Zs \Omega_0} \subset E_0 \Hs$. We shall see that 
the converse also holds. But, first, we adapt classic arguments 
\cite{ReSch,Bor1,Ar1,DS} to prepare some intermediate results. 
Let $a \in \RR^4$ be a spacelike translation, and set $a_n = na$, 
for each $n \in \IN$. Let $\Os \in \Cs$, $A \in \Rs(\Os)$ and 
$A(a_n) = U(a_n) A U(a_n)^{-1}$. Since the sequence $\{ A(a_n) \}$ 
is uniformly bounded in norm and $\Hs$ is separable, there exists 
a subsequence $\{ A(a_{n_k}) \}$ which is weakly convergent. By 
the standing assumption on $\Cs$, for any $\widetilde{\Os} \in \Cs$ 
there exists an $N \in \IN$ and 
a wedge $W_N \in \Ws$ such that $\widetilde{\Os} \subset W_N$ 
and, for every $n \geq N$,  
$\Os + na \subset W_N{}'$, \ie 
$A(a_n) \in \Rs(W_N{}') = \Rs(W_N)' \subset \Rs(\widetilde{\Os})'$. 
Since $\Rs$ is generated by the algebras $\Rs(\widetilde{\Os})$, 
$\widetilde{\Os} \in \Cs$, the weak limit of the corresponding 
subsequence $\{ A(a_{n_k}) \}$, call it $A_{\infty}$, is an 
element of $\Rs' = \Zs$.  Moreover, 
\cite[Lemma 4]{Bor1} implies
\begin{equation}  \label{infinity}
A_{\infty} \Omega_0 = w-\lim_{k \rightarrow \infty} A(a_{n_k}) \Omega_0 =
w-\lim_{k \rightarrow \infty} U(a_{n_k}) A \Omega_0 = E_0 A \Omega_0 \, .
\end{equation}
Let $\Ys = \{ A_{\infty} \mid A \in \Rs(\Os), \Os \in \Cs \} \subset \Zs$ 
denote the set of all such weak limit points. Since 
$\Omega_0$ is cyclic for  $\Rs$
it follows from relation (\ref{infinity}) that $\Ys \Omega_0 $
is a dense subset of $E_0 \Hs$. Thus, since $\Ys \subset \Zs$
and $\Zs \Omega_0 \subset E_0 \Hs$ we arrive at the following statement.
\begin{prop}  \label{vacuum}
Under the standing assumptions, one has 
$E_0 \Hs = \overline{\Zs\Omega_0} = \overline{\Ys\Omega_0}$.
\end{prop}

     The following result is an easy consequence of 
the preceding proposition and the inclusion $\Ys \subset \Zs$,
established before.

\begin{cor} Given the standing assumptions, one has the equality
$\Ys = \Zs$.
\end{cor}

\begin{proof}
It was shown in Proposition \ref{vacuum} that 
$ \overline{\Ys\Omega_0} = E_0 \Hs $.
Thus, the restriction of the abelian algebra 
$\Ys$ to the subspace $E_0 \Hs$ has $\Omega_0$ as 
a cyclic vector. It follows that $\Ys$ 
is maximally abelian on $E_0 \Hs$. Since $\Ys$ is contained in the abelian
algebra $\Zs$, the restrictions of $\Ys$ and $\Zs$ to $E_0
\Hs$ must coincide.  The desired assertion then follows, because
$\Omega_0$ is separating for $\Zs$.
\end{proof}

  In a vacuum representation fulfilling the standard assumptions,
including the spectrum condition, it is known \cite{DS} that $\Zs =
\Zs(W)$, for all $W \in \Ws$, but this need not be the case in the
setting considered here.

     After these preparations, we proceed to the central decomposition
of $\Rs$. Since the center $\Zs$ of $\Rs$
coincides with the commutant $\Rs^\prime$, this amounts
to a decomposition of the underlying Hilbert space into 
irreducible subsectors. Moreover, as the Poincar\'e transformations
are weakly inner, the  representation $U(\Ps_+)$ decomposes into 
a continuous unitary representation of
$\Ps_+$ in each sector. In particular, the Lorentz group is not
spontaneously broken by this decomposition, which is to be contrasted
with the existence of examples of nets in vacuum representations
(satisfying the spectrum condition but not the CGMA) in which the
Lorentz group is spontaneously broken in the central decomposition of
$\Rs$ \cite{Ar1,DS}.  In \cite{DS} it was shown that modular
covariance (see below for a definition) prevents spontaneous breaking
of the Lorentz group; here it is the CGMA which assures the 
stability of each vacuum sector under the action of the Lorentz
group. Note that the CGMA is known to hold more generally than
modular covariance does \cite{BDFS}.

     The proof of our decomposition theorem rests upon the theory of
direct integral decomposition of a von Neumann algebra presented in
\cite{Dix}. The algebra $\Rs$ is decomposed with respect to the
abelian algebra $\Zs$ to yield a standard Borel measure
space $(S,\nu)$ and measurable families $\zeta \mapsto \Hs(\zeta)$ of
Hilbert spaces and $\zeta \mapsto \Rs(\zeta)$ of von Neumann algebras
such that
$$\Hs = \int_S^\oplus \Hs(\zeta) \; d\nu(\zeta) \, , \, 
\Rs = \int_S^\oplus \Rs(\zeta) \; d\nu(\zeta) \, . $$
For $\nu$-almost all $\zeta$, $\Rs(\zeta)$ is a factor 
\cite[Thm.\ II.3.3]{Dix}. 

     But here we are concerned with the decomposition of a great deal
more structure. Though it is clear from Proposition  \ref{centerinvariant}
and \ref{vacuum} that the algebras $\Rs(W)$, $W \in \Ws$, and the
group $U(\Ps_+) = \Js$ also decompose, it is necessary to find 
a set $N \subset S$ with
$\nu(N) = 0$ such that for every $\zeta \in S \setminus N$ all of the
decomposed structures still have the original properties. However,
this involves {\it prima facie} uncountably many conditions, which
could lead to a zero-set catastrophe.

     The standard technique to handle this technical problem is to impose
only countably many of these conditions, each of which would hold for
all $\zeta$ except in a set of measure zero. Since $\nu$ is countably
additive, all countably many conditions would hold except in a
possibly larger set $N$ of measure zero. One then employs a suitable
limit argument to assure that the remaining conditions also hold for
all $\zeta \in S\setminus N$.  Of course, a countable union of
countable sets is countable, and it is only a matter of taste or
convenience whether one imposes in the argument the countable union of
conditions at once or each countable subset after the other. Since the
decomposition of many of the structures we are concerned with here has
already been carefully treated in the literature, we shall only
indicate details which seem to involve new arguments.

     We recall some facts from \cite{BDFS,BS2}. Making use of the fact
that $\Pid$ acts transitively on $\Ws$, we identify $\Ws$, as a
topological space, with the quotient space $\Pid / \Ps^{}_0 $, where
$\Ps^{}_0 \subset \Pid$ is the invariance subgroup of any given wedge
$W_0 \in \Ws$; note that the topology does not depend on the choice of
$W_0$. As $\Pid / \Ps^{}_0 $ is separable, so is $\Ws$. In order
to successfully decompose all the structures of interest in such a
manner that the zero set catastrophe is avoided, we need to be 
able to choose a countable, dense 
subgroup $\widehat{\Ps} \subset \Pid$ and a countable, dense subset 
$\widehat{\Ws} \subset \Ws$ satisfying the following conditions:
\begin{enumerate}
\item[(i)] The elements of $\widehat{\Ps} $ 
leave $\widehat{\Ws}$ stable 
and $\widehat{\Ps} $ acts transitively upon $\widehat{\Ws}$. 
\item[(ii)] For any $W_1,W_2 \in \Ws$ such that $W_1 \subset W_2$, 
there exist two sequences \newline
$\{ W_{1,n}\},\{ W_{2,n}\} \subset \widehat{\Ws}$ 
such that $\{ W_{i,n}\}$ coverges to $W_i$, $i = 1,2$, and \newline
$W_{1,n} \subset W_{2,n}$, for all $n \in \NN$.
\end{enumerate}

\noindent The reader may verify that these conditions are fulfilled if
$\widehat{\Ps}$ is chosen to be the semi--direct product 
of rational translations with the image under the canonical projection 
homomorphism of the subgroup of the covering group 
\textit{SL(2,$\CC$)} whose elements have entries with only rational real and
imaginary parts, and $\widehat{\Ws}$ is chosen to be $\widehat{\Ps} \, W_0$ 
for some fixed wedge $W_0$. 

\begin{theorem} \label{decomposition}
Under the standing assumptions, the central decomposition of $\Rs$
leads to a unique\footnote{The measure space $(S,\nu)$ is unique up 
to isomorphism, and given $(S,\nu)$ the measurable fields are unique 
up to unitary equivalence. See Section II.6.3 in \cite{Dix} for details.}
integral decomposition of the given structures into
irreducible, Poincar\'e-covariant nets. Precisely, there exists a 
measure $\nu$ on the spectrum $S$ of $\Zs$ and measurable families
of Hilbert spaces $\zeta \rightarrow \Hs(\zeta)$, von Neumann algebras
$\zeta \rightarrow \Rs(\zeta) \subset \Bs(\Hs(\zeta))$, and strongly
continuous (anti)unitary representations of the proper Poincar\'e
group $\zeta \rightarrow U(\Ps_+)(\zeta)$ such that
$$\Hs = \int_S^\oplus \Hs(\zeta) \; d\nu(\zeta) \, , \, 
\Rs = \int_S^\oplus \Rs(\zeta) \; d\nu(\zeta) \, , \,
U(\lambda) = \int_S^\oplus U(\lambda)(\zeta) \; d\nu(\zeta) \, , $$
for all $\lambda \in \Ps_+$. Moreover, for each $W \in \Ws$, there
exists a measurable family of von Neumann algebras 
$\zeta \rightarrow \Rs(W)(\zeta) \subset \Bs(\Hs(\zeta))$ such that
\begin{equation} \label{wedgealgebra}
\Rs(W) = \int_S^\oplus \Rs(W)(\zeta) \; d\nu(\zeta) \, , 
\end{equation}
and such that isotony is satisfied by $\{\Rs(W)(\zeta)\}_{W \in \Ws}$
$\nu$-almost everywhere. For $\nu$-almost all $\zeta$, 
$\Rs(\zeta) = \Bs(\Hs(\zeta))$, $E_0(\zeta)\Hs(\zeta) = (E_0 \Hs)(\zeta)$ 
is one-dimensional, and 
\begin{equation} \label{covar}
U(\lambda)(\zeta) \; \Rs(W)(\zeta) \; U(\lambda)(\zeta)^{-1} =
\Rs(\lambda W)(\zeta) \, ,
\end{equation}
for all $\lambda \in \Ps_+$ and $W \in \Ws$.  
\end{theorem}

\begin{proof}
The decomposition of the Hilbert space and algebra $\Rs$ is explained
in \cite{Dix}. As already mentioned, the factorial components 
$\Rs(\zeta)$ in the central decomposition of $\Rs$ indeed act 
irreducibly on the respective subspaces $\Hs(\zeta)$ since $\Zs = \Rs^\prime$
by Proposition~\ref{typeone}.
The representation $U(\Pid)$ of the identity component 
of the Poincar\'e group and the subspace $E_0 \Hs$ are decomposed 
in \cite{DS}, and the attendant assertions made above
are proven there, using results in \cite{Ju}. Although the net
$\wrnet$ was also decomposed in \cite{DS}, there the argument was 
framed for locally generated nets for which $\Cs$ is the set of 
double cones; a concrete choice of a countable ``dense'' subcollection 
of double cone algebras was given there. To obtain the assertion in
the generality made here, one must provide another argument.

     Instead, here one decomposes the elements of the countable set 
$\{ \Rs(W) \}_{W \in \widehat{\Ws}}$ to obtain for each 
$W \in \widehat{\Ws}$ a measurable family 
$\zeta \mapsto \Rs(W)(\zeta)$ such that (\ref{wedgealgebra})
holds. By enlarging the zero set $N$, if necessary, the covariance 
(\ref{covar}) in $\nu$-almost all sectors holds for all
$W \in \widehat{\Ws}$ and $\lambda \in \widehat{\Ps}$. 

     Theorem II.3.1 in \cite{Dix} guarantees that for a fixed pair of
wedges such that $W_1 \subset W_2$, the containment 
$\Rs(W_1)(\zeta) \subset \Rs(W_2)(\zeta)$ holds $\nu$-almost
everywhere. After a possible change of the set
$N$, the same is true for all $W_1,W_2 \in \widehat{\Ws}$ with 
$W_1 \subset W_2$. 

     For an arbitrary $W \in \Ws$, there exists an element 
$\lambda_0 \in \Pid$ such that $W = \lambda_0 W_0$. By construction,
there exists a sequence $\{\lambda_n \} \subset \widehat{\Ps}$ 
which converges to $\lambda_0$. Define 
$$ \Rs(W)(\zeta) = \{ w-\lim_{n \rightarrow \infty} 
U(\lambda_n)(\zeta) A(\zeta) U(\lambda_n)(\zeta)^{-1} \mid A(\zeta)
\in \Rs(W_0)(\zeta) \} \, .$$
The strong continuity of $U(\Pid)(\zeta)$ in these sectors entails
that $\Rs(W)(\zeta)$ is independent of the choice of such a sequence.
Moreover, the same continuity implies that (\ref{covar}) is valid for all 
$W \in \Ws$, $\lambda \in \Pid$, and the definition of 
$\Rs(W)(\zeta)$ is compatible with all elements of the construction. 
In particular 
$\Rs(W)(\zeta) =  
U(\lambda_0)(\zeta) \Rs(W_0)(\zeta) U(\lambda_0)(\zeta)^{-1}$. 
By the measurability of 
$\zeta \mapsto U(\Pid)(\zeta)$ and the 
covariance of the original net, it follows that the family 
$\zeta \mapsto \Rs(W)(\zeta)$ is measurable and that
(\ref{wedgealgebra}) holds for all $W \in \Ws$. The isotony in 
$\nu$-almost all sectors for wedge algebras indexed by 
the elements of $\Ws$ now
follows easily from property (ii) above and the already-established 
isotony for wedge algebras indexed by $\widehat{\Ws}$.

     Finally, as $U(\Ps_+^\downarrow) = U(\lambda_W)U(\Pid)$, for fixed 
$W \in \Ws$, the assertion concerning $U(\Ps_+)$ follows, since the 
complex antilinearity of $U(\lambda_W) = J_W$, \ie the fact that 
$U(\lambda_W)$ commutes with $\Zs_{s}$ but not with $\Zs$, poses 
no problems \cite[Thm.~III.2]{Ju}.
\end{proof}

   It is noteworthy that, in our general setting, the above central 
decomposition always results in irreducible sectors even though 
the spectrum condition need not hold. This is in contrast to the 
situation in the Wightman formalism where the extremal states 
resulting from a corresponding decomposition need not be pure
states \cite{Bor2,BoYn} -- cf.\ \cite{DS2} for a discussion of this
matter.

    We close this section with a comment about unbroken symmetries
in the internal symmetry group $G$. The group $G$ will be unitarily
implemented in a given sector if and only if $V(G)$ commutes with
the corresponding projection in $\Zs$. On the other hand,
if $G$ is a separable topological group, the
representation $g \mapsto V(g)$ is strongly continuous, and there
exists a subgroup $H \subset G$ such that $\Zs \subset V(H)'$,
then the above arguments entail that there
exists a measurable family of strongly continuous unitary representations
$\zeta \rightarrow V(H)(\zeta)$ such that
$$V(h) = \int_S^\oplus V(h)(\zeta) \; d\nu(\zeta) \, , $$
and 
$$V(h)(\zeta) \; \Rs(W)(\zeta) \; V(h)(\zeta)^{-1} = \Rs(W)(\zeta) \, ,$$
for all $h \in H$, $W \in \Ws$ and $\nu$-almost all $\zeta$.

   In the next section, we prove that the modular structure
associated with pairs $(\Rs(W),\Omega_0)$, $W \in \Ws$, also decomposes
in such a manner that conditions (a)--(c) are satisfied in
$\nu$-almost all sectors.

\section{The rigidity of geometric modular action} 
\setcounter{equation}{0}

    We maintain the standing assumptions in this section and turn our
attention to the modular structures, their properties and their
behavior under the central decomposition carried out above.  Let
\begin{equation} \label{natural}
\Ps_W^\natural = \overline{\{ \Delta_W^{1/4}A \Omega_0 \mid A \in \Rs(W)_+ \}}
\end{equation}
denote the natural positive cone corresponding to the pair $(\Rs(W),\Omega_0)$, 
where $\Rs(W)_+$ is the set of all positive elements in $\Rs(W)$,
and let
$$\Ps_0 = {\bigcap}_{\, W \in \Ws} \; \Ps_W^\natural . $$
Of course, we have $\Omega_0 \in \Ps_0$. As shown in \cite{Ar2}, every
vector $\Phi \in \Ps_W^\natural$, which is either cyclic or separating
for $\Rs(W)$, is both cyclic and separating for $\Rs(W)$. Moreover,
the modular conjugation $J_W^\Phi$ corresponding to the pair
$(\Rs(W),\Phi)$ coincides with $J_W$ \cite[Thm.\ 4]{Ar2}. Hence, if 
$\Omega \in \Ps_0$ is cyclic or separating for all $\Rs(W)$, $W \in \Ws$, 
then $J_W^\Omega = J_W$, for every $W \in \Ws$. Thus, the pair 
$(\wrnet,\Omega)$ must also fulfill conditions (a)--(c), if 
$(\wrnet,\Omega_0)$ does. The CGMA therefore selects state vectors which 
lie in $\Ps_0$, and so we wish to investigate the structure of $\Ps_0$ and 
the properties of the states determined by the elements of $\Ps_0$.
 
     We begin with the following lemma.

\begin{lemma}  \label{cone}
Under the standing assumptions, $\Ps_0$ is a pointed,
weakly closed convex cone such that 
\begin{equation} \label{positivity}
\langle\Omega, A J_W A \Omega_0\rangle \geq 0 \, ,
\end{equation}
for all $\Omega \in \Ps_0$, $W \in \Ws$ and $A \in \Rs(W)$.
\end{lemma}

\begin{proof}
It is shown in \cite[Thm.\ 4]{Ar2} that $\Ps_W^\natural$ is a pointed,
weakly closed, selfdual convex cone. Since $\Ps_0$ is an intersection of
these cones, it is clearly a weakly closed convex cone.  Moreover, if
$\Omega$ and $-\Omega$ are contained in $\Ps_0$, they are also in
$\Ps_W^\natural$; hence, $\Omega = 0$. In the same theorem it is shown
that $\langle\Omega, A J_W A \Omega_0\rangle \geq 0$, for all
$\Omega \in \Ps_W^\natural$ and $A \in \Rs(W)$. Since
$\Ps_0 \subset \Ps_W^\natural$, for all $W \in \Ws$, the final assertion
follows.
\end{proof}

     This lemma enables us to prove the following result.

\begin{prop} \label{invariance}
Under the standing assumptions, every element of $\Ps_0$ is
invariant under $U(\Ps_+)$; in particular, $\Ps_0 \subset E_0 \Hs$.
In fact, $E_0 \Hs$ is the linear span of $\Ps_0$ and 
$\Ps_0 = \overline{\Zs_+ \Omega_0}$. 
\end{prop}

\begin{proof}
Theorem 4 (2) in \cite{Ar2} entails that if $\Omega \in \Ps_0$, then $J_W
\Omega = \Omega$, for all $W \in \Ws$. Since $U(\Ps_+) = \Js$, one has
$U(\lambda)\Omega = \Omega$, for every $\lambda \in \Ps_+$.  
Thus, in particular, $\Ps_0 \subset E_0 \Hs$. A basic
result of modular theory (cf. \cite[Lemma 3.2.16]{BrRo}) entails 
that every element of the center
$\Zs(W)$ is left invariant by the adjoint action of the modular
unitaries $\Delta_W^{it}$, $t \in \RR$. Hence,
Proposition \ref{centerinvariant} implies that for every $Z \in \Zs_+$
one has $\Delta_W^{1/4} Z \Omega_0 = Z \Omega_0$, and thus $Z \Omega_0 \in
\Ps_W^\natural$, for every $W \in \Ws$, by (\ref{natural}). This
entails the inclusion $\Zs_+ \Omega_0 \subset \Ps_0$. Proposition \ref{vacuum}
then implies that $E_0 \Hs$ is the linear span of $\Ps_0$. 

     Since $\Ps_0 \subset E_0 \Hs = \overline{\Zs\Omega_0}$,
there exists a normal operator $Z$ affiliated with $\Zs$ such
that $\Omega = Z\Omega_0$. But for any $A \in \Zs_+$ one has
$A = A^{1/2}J_W A^{1/2}J_W$, so that (\ref{positivity}) yields
$\langle\Omega,A\Omega_0\rangle \geq 0$, for all $A \in \Zs_+$.
Setting $A = B^* B$, $B \in \Zs$, this implies
$$0 \leq \langle Z \Omega_0, B^* B \Omega_0\rangle =  
 \langle Z B \Omega_0, B \Omega_0\rangle \, .$$
The restriction of $Z$ to $\Zs\Omega_0$ is therefore positive.
But $Z$ can be decomposed into four positive operators 
$Z_+,Z_-,\widetilde{Z}_+,\widetilde{Z}_-$ affiliated with $\Zs$ such that
$Z = Z_+ - Z_- + i(\widetilde{Z}_+ - \widetilde{Z}_-)$, and since $\Omega_0$
is separating for $\Zs$, it follows that $Z = Z_+$.
\end{proof}

     Although the modular conjugations associated with a given von
Neumann algebra and different cyclic and separating vectors from
$\Ps^\natural$ coincide, typically the corresponding modular unitaries
differ from vector to vector. However, the rigidity of
the structure investigated here carries through also to the 
modular operators.

\begin{cor} \label{samemodularoperator}
Under the standing assumptions, if $\Omega \in \Ps_0$ is cyclic or separating
for $\Rs(W)$ and $\Delta_W^\Omega$ is the associated modular operator,
then $\Delta_W^\Omega = \Delta_W$.
\end{cor}

\begin{proof}
By Proposition \ref{invariance}, there exists a positive operator $Z$ 
affiliated with $\Zs$ such that $\Omega = Z \Omega_0$. This operator, 
just as every
positive element of $\Zs \subset \Zs(W)$, commutes with 
the antiunitary $J_W$, the algebra $\Rs(W) \bigvee \Rs(W)'$ and 
with any modular group associated with $\Rs(W)$. Hence, 
for any $A \in \Rs(W)$ one has
\begin{eqnarray*}
(\Delta_W^\Omega)^{1/2}A\Omega & = & J_W A^* \Omega = J_W A^* Z \Omega_0 = 
J_W A^* J_W Z \Omega_0 \\
& = & Z J_W A^* \Omega_0 = Z \Delta_W^{1/2} A\Omega_0 = \Delta_W^{1/2} ZA \Omega_0 \\
& = & \Delta_W^{1/2} AZ \Omega_0 = \Delta_W^{1/2} A \Omega \, ,
\end{eqnarray*}
where $J_W \Rs(W) J_W = \Rs(W)'$ has also been used. Thus, one 
concludes $\Delta_W^\Omega \subset \Delta_W$. A similar argument
interchanging the roles of $\Omega$ and $\Omega_0$ completes the proof.
\end{proof}

     In light of the fact that any normal state on $\Bs(\Hs)$, when
restricted to $\Rs(W)$, can be implemented on $\Rs(W)$ by a suitable
vector in $\Ps_W^\natural$ \cite[Theorem 6]{Ar2}, it is noteworthy
that these different implementers sit in the various natural positive
cones $\Ps_W^\natural$ in such a way that only very well-behaved
states are determined by the vectors left in the intersection $\Ps_0$.

     Proposition \ref{invariance} also entails that under the central
decomposition of $\Rs$, in $\nu$-almost all $\Hs(\zeta)$ the
corresponding set $\Ps_0(\zeta)$ contains only vectors proportional to
$\Omega_0(\zeta)$. Hence, in each irreducible vacuum sector at most 
one state can satisfy the CGMA in the form of conditions (a)--(c).

     The next theorem establishes the properties under central 
decomposition of the various modular structures of concern to us.

\begin{theorem}  \label{modulardecomp}
Under the standing assumptions, in reference to the structures discussed
in Theorem \ref{decomposition}, let, for each $W \in \Ws$,
$J_W(\zeta),\Delta_W(\zeta),\Ps_W^\natural(\zeta)$ represent the modular 
objects associated with the pair $(\Rs(W)(\zeta),\Omega_0(\zeta))$, where
$\Omega_0 = \int_S^\oplus \Omega_0(\zeta) \; d\nu(\zeta)$. Then for
each $W \in \Ws, t \in \RR$, the fields $\zeta \mapsto J_W(\zeta)$,
$\zeta \mapsto \Delta_W^{it}(\zeta)$ and 
$\zeta \mapsto \Ps_W^\natural(\zeta)$ are measurable and
$$J_W = \int_S^\oplus J_W(\zeta) \; d\nu(\zeta) \, , \,
 \Delta_W^{it} = \int_S^\oplus \Delta_W^{it}(\zeta) \; d\nu(\zeta) \, , \,
 \Ps_W^\natural = \int_S^\oplus \Ps_W^\natural(\zeta) \; d\nu(\zeta) \, .$$
Conditions (a)--(c) hold in $\nu$-almost all sectors.
If, moreover, $\Ps_0(\zeta) = \bigcap_{W \in \Ws}\Ps_W^\natural(\zeta)$,
then also $\zeta \mapsto \Ps_0(\zeta)$ is measurable and
$$\Ps_0 = \int_S^\oplus \Ps_0(\zeta) \; d\nu(\zeta) \, .$$
For $\nu$-almost all $\zeta$, 
$\Ps_0(\zeta) = \{ c \, \Omega_0(\zeta) \mid c \in [0,\infty) \}$.
\end{theorem}

\begin{proof}
For every $W \in {\Ws}$, the measurability of the fields 
$\zeta \mapsto J_W(\zeta)$, $\zeta \mapsto \Delta_W^{it}(\zeta)$
and the equalities
$$J_W = \int_S^\oplus J_W(\zeta) \; d\nu(\zeta) \, , \,
 \Delta_W^{it} = \int_S^\oplus \Delta_W^{it}(\zeta) \; d\nu(\zeta)$$
are assured by \cite[Thm.\ III.2]{Ju}. {}From Theorem \ref{decomposition} 
it follows that
$$ J_W = U(\lambda_W) = \int_S^\oplus U(\lambda_W)(\zeta) \; d\nu(\zeta)\, , $$
for every $W \in \Ws$. Corollary II.2.2 in \cite{Dix} then yields the
equality $J_W(\zeta) = U(\lambda_W)(\zeta)$ for $\nu$-almost all $\zeta$.
With a possible change in the zero set $N$, this equality may be assured 
for all $W \in \widehat{\Ws}$. In Section 3 of \cite{BS2} it was
shown that for a locally generated net satisfying Haag duality the map 
$W \mapsto J_W$ from the space of wedges to the topological group
$\Js$ is continuous, as is the map $\lambda_W \mapsto J_W$ from
$\Ps_+$ to $\Js$. This continuity and the continuity of the 
representation $U(\Ps_+)$ entail then that condition (c) holds 
in $\nu$-almost all sectors.

     The isotony in $\nu$-almost every sector was established in 
Theorem \ref{decomposition}. From \cite[Prop. II.2]{Ju} it follows 
that for a fixed $W_0 \in \Ws$, $\Omega_0(\zeta)$ is cyclic and 
separating for $\Rs(W_0)(\zeta)$ for $\nu$-almost all $\zeta$. In 
view of the covariant action of the unitaries $U(\lambda)(\zeta)$ 
on the wedge algebras $\Rs(W)(\zeta)$ proven in Theorem 
\ref{decomposition} and the transitivity of $\Pid$ on $\Ws$, this 
is therefore true for all $W \in \Ws$ and the same set of $\zeta$. 
Hence, conditions (a)--(c) with the possible exception of the 
injectivity in (a) hold in $\nu$--almost all sectors.

     By Proposition A.1, if the map $W \mapsto \Rs(W)(\zeta)$ is not injective,
then $U(\Pid)(\zeta)$ is trivial and $\Rs(W)(\zeta)$ is abelian and
independent of $W \in \Ws$. But then $\Rs(\zeta)$ is an abelian factor
with cyclic vector. If this were true for all $\zeta$ in a measurable
set $M \subset S$ with positive $\nu$-measure, then $\int_M^\oplus
\Hs(\zeta) \; d\nu(\zeta)$ would be a subspace of $\Hs$ on which the
corresponding subrepresentation of $U(\Pid)$ was trivial and of $\Rs$
was abelian. This degenerate situation has been excluded by the
standing assumptions.

     Since $\Delta_W$ commutes with $\Zs(W)$, 
Proposition \ref{centerinvariant} implies that it commutes 
with $\Zs$ and hence is also decomposable.
Appealing to \cite[Thm.\ I.8, Thm.\ III.2]{Ju}, it follows that 
$$ \Delta_W^{1/4} A \Omega_0 = 
 \int_S^\oplus \Delta_W^{1/4}(\zeta) A(\zeta) \Omega_0(\zeta) \; d\nu(\zeta)\, , 
$$
for all $A \in \Rs(W)$, and therefore that
$$ \Ps_W^\natural = \int_S^\oplus \Ps_W^\natural(\zeta) \; d\nu(\zeta) \, ,$$
for all $W \in \Ws$.

     As the standing assumptions hold in $\nu$-almost all sectors,
one may apply Proposition \ref{invariance} in each sector to conclude
$\Ps_0(\zeta) = \overline{\Zs(\zeta)_+ \Omega_0} = 
\overline{\Zs_+(\zeta) \Omega_0}$, and thereby also 
$\Ps_0 = \int_S^\oplus \Ps_0(\zeta) \; d\nu(\zeta)$. Since for $\nu$-almost
all $\zeta$ the elements of $\Zs_+(\zeta)$ are positive multiples
of the identity operator on $\Hs(\zeta)$, the final assertion is
immediate. 
\end{proof}

     We remark that also all of the modular unitaries 
$\{ \Delta_W^{it}\}_{t \in \RR}$ can be reunited in $\nu$-almost all
sectors as above by first decomposing the operators $\Delta_W^{it}$,
for $t$ rational and $W \in \widehat{\Ws}$, and then using the 
strong continuity to reconstruct $\Delta_W^{it}(\zeta)$ for all 
$t \in \RR$. From Section 3 of \cite{BS2} and \cite[Prop. 4.6]{BDFS} 
one knows that also the map $W \mapsto \Delta_W^{it}$ is strongly 
continuous, given our standing assumptions. This is then employed 
to reconstruct $\Delta_W^{it}(\zeta)$ for all $W \in \Ws$.

     A conceptually simple and quite general criterion for stable
states on general space--times is the \msc, proposed in \cite{BDFS}.
We recall this condition here for the convenience of the reader.
\begin{enumerate}
\item[(d)] For any $W \in \Ws$, the elements $\Delta_W^{it}$, $t \in \RR$, 
of the modular group corresponding to $(\Rs(W), \Omega_0)$ are contained 
in the group $\Js$ generated by all finite products of the modular
involutions $\{ J_W\}_{W \in \Ws}$.
\end{enumerate}
We refer the interested reader to \cite{BFS,BDFS} for a discussion of
the background of this condition and a brief account of other
interesting approaches towards an algebraic characterization of ground
states on general space--times. As shown in \cite{BDFS}, if the
standing assumptions of this paper and the \msc\, hold, then modular
covariance obtains: $\Delta_W^{it} = U(\lambda_W(2\pi t))$, for all $t
\in \RR$ and $W \in \Ws$, where $\{\lambda_W(2\pi t) \mid t \in\RR\}$
is the one-parameter subgroup of boosts leaving $W$ invariant. In
addition, the spectrum condition holds.

     We close this section with a theorem which summarizes the 
consequences of the \msc\, for the topics under consideration
here.  

\begin{theorem}
If the standing assumptions and the \msc\, hold for $\Omega_0$, then the
conditions (a)--(d) also obtain for any $\Omega \in \Ps_0$
which is cyclic or separating for all wedge algebras $\Rs(W)$. In
addition, $\Rs' = \Zs = \Zs(W)$, for every $W \in \Ws$. 
Hence, the central decomposition in Theorem \ref{decomposition}
results in irreducible vacuum sectors in which the \msc\, is satisfied
in $\nu$-almost every sector, as is modular covariance and the
spectrum condition. In $\nu$-almost all sectors, $\Rs(W)(\zeta)$
is a type III$_1$ factor, for all $W \in \Ws$.
\end{theorem}

\begin{proof}
It has already been shown that conditions (a)--(c) hold for every
$\Omega$ as described. Corollary \ref{samemodularoperator} entails that
also condition (d) is satisfied by the modular unitaries associated to
each wedge algebra by such vectors $\Omega$ (and, of course, they
manifest modular covariance). Together, Proposition 5.1 and the proof of
Theorem 5.1 in \cite{BDFS} entail that $U(\RR^4)$ fulfills the
spectrum condition. It then follows from \cite[Prop.\ 3.1]{DS} that
$\Zs(W) = \Zs = \Rs'$, for every $W \in \Ws$.  Therefore, in
the central decomposition in Theorem \ref{decomposition} 
one has the spectrum condition for
$U(\RR^4)(\zeta)$, for $\nu$-almost all $\zeta$ \cite[Thm.\ 4.1]{DS}.
Furthermore, from the proof of Lemma 3.2 in \cite{DS} one may conclude
that $\Rs(W)(\zeta)$ is a type III$_1$ factor, for all $W \in \Ws$.

     {}From the proof of Theorem \ref{modulardecomp} and
\cite[Cor.\ II.2.2]{Dix}, it follows that for $\nu$-almost all $\zeta$
one has $\Delta_W^{it}(\zeta) = U(\lambda_W(2\pi t))(\zeta)$, for all
$t \in \RR$ and $W \in \Ws$, \ie modular covariance holds in
$\nu$-almost all sectors. {}From the proof of Theorem 
\ref{modulardecomp} it also follows that 
$U(\Ps_+)(\zeta) = \Js(\zeta)$, for $\nu$-almost all $\zeta$.
Therefore, the \msc\, also holds in $\nu$-almost all sectors.
\end{proof}

     We mention that if the hypothesis of Theorem 4.5 holds, then
one can show using \cite[Thm.\ 5.1]{BDFS} and \cite[Thm.\ 1.2]{PuWo}
that any $\Omega \in \Ps_0$ and the corresponding $\Omega(\zeta)$, 
for $\nu$-almost all $\zeta$, determine passive states on their 
respective nets with respect to all uniformly accelerated observers. 
Hence, the CGMA and the \msc\, select particularly stable states. 

\section{Final comments}

     A number of different criteria \cite{FH,KaWa,Ra,Ra2,BFK} 
have been proposed to select physically relevant states for
quantum field theories on curved space--times, where translation
covariance and the spectrum condition are simply not
applicable. However, these criteria, when they obtain, are valid for
an entire folium of states and therefore beg the question of which
state (or states) of the respective folium is to be regarded as
fundamental, \ie as a reference or ground state \cite{BDFS}.

     We emphasized in \cite{BDFS} that the CGMA is a selection
criterion for states and not an entire folium. However, the CGMA
explicitly places constraints only on the algebras $\Rs(W)$, $W \in
\Ws$, and the modular conjugations $J_W$, $W \in \Ws$ --- the algebras
are state-independent and each modular conjugation $J_W$ is common to
every state vector in the natural cone $\Ps_W^\natural$, which is
itself so large that it spans the Hilbert space $\Hs$. However, the
CGMA is a condition on the entire set $\{ J_W \mid W \in \Ws \}$, and
therefore the vectors selected by the CGMA are those in $\Ps_0$. 

     We have shown in this paper that the vectors remaining in the
intersection $\Ps_0$ share the properties one would desire of reference
states, without any appeal to the spectrum condition, and that the
structures associated with the CGMA and the \msc\, are gratifyingly
rigid.  Moreover, we have shown that these conclusions do not rely
upon the more technical assumptions of the CGMA in Minkowski space
\cite{BDFS}, which were designed to assure not only the existence of
the representation of the Poincar\'e group discussed above, but also
to derive the Poincar\'e group and its action upon Minkowski space
from the initial data $(\wrnet,\Omega_0)$.  Already the conditions
(a)--(c), themselves consequences of the CGMA in Minkowski space, are
sufficient to assure the above-mentioned conclusions.

\bigskip

\noindent {\bf Acknowledgements}:
We are grateful to the anonymous referee for 
drawing our attention to Proposition \ref{typeone}, which 
simplified the discussion.  
DB wishes to thank the Institute for Fundamental Theory and the
Department of Mathematics of the University
of Florida and SJS wishes to thank the Institute
for Theoretical Physics of the University of G\"ottingen 
for hospitality and financial support which facilitated this research.
This work was supported in part by a research grant of 
Deutsche Forschungsgemeinschaft (DFG). 

\appendix

\section{Nets of wedge algebras}

     We show that in the presence of the other standing assumptions,
the injectivity of the map $W \mapsto \Rs(W)$ can fail only in the
most extreme manner.

\begin{prop}  \label{bijection}
Let all of the standing assumptions hold, except condition (a). If
the map $W \mapsto \Rs(W)$ is order-preserving but
not injective, then the representation $U(\Pid)$ is trivial, and
$\Rs(W)$ is abelian and independent of $W \in \Ws$.
\end{prop}

\begin{proof}
Let $W_1,W_2 \in \Ws$ be distinct wedges such that $\Rs(W_1) = \Rs(W_2)$.
Then the corresponding modular conjugations must coincide, \ie
$J_{W_1} = J_{W_2}$. Since $W_1 \neq W_2$, condition (c) and 
$U(\Ps_+) = \Js$ entail the existence of a nontrivial element
$\lambda_0 = \lambda_{W_1}\lambda_{W_2} \in \Pid$ such that 
$U(\lambda_0) = J_{W_1}J_{W_2} = 1$.
With $\lambda_0 = (\Lambda_0,x_0)$, $\Lambda_0 \in \Lid$, $x_0 \in \RR^4$,
one would then have 
$$U(x_0)^{-1} = U(x_0)^{-1} U(\lambda_0) = U(\Lambda_0) \, .$$ 
Hence, $U(\Lambda_0) \in U(\RR^4)$, and since $U(\RR^4)$ is a normal
subgroup of $U(\Pid)$ it then follows that 
$$U(\Lambda \Lambda_0 \Lambda^{-1}) = U(\Lambda)U(\Lambda_0)U(\Lambda)^{-1}
\in U(\RR^4) \, ,$$
for every $\Lambda \in \Lid$. The elements 
$\{ \Lambda \Lambda_0 \Lambda^{-1} \mid \Lambda \in \Lid \}$ generate
a (nontrivial) normal subgroup of $\Lid$. But $\Lid$ is a simple group, so
one deduces that $U(\Lid) \subset U(\RR^4)$. The representation 
$U(\Lid)$ is therefore abelian and hence trivial. But then for every
$x \in \RR^4$ and $\Lambda \in \Lid$ one has 
$U(x) = U(\Lambda)U(x)U(\Lambda)^{-1} = U(\Lambda x)$, so that it follows
that $U(x)$ is independent of $x \in \RR^4$. Thus, $U(\Pid)$ is
trivial. But the covariance of the net under the adjoint action of 
$U(\Pid)$ then entails that $\Rs(W) = \Rs(\lambda W)$, for all 
$\lambda \in \Pid$. Thus, one must conclude, in particular, that 
$\Rs(W) = \Rs(\lambda_W W) = \Rs(W') = \Rs(W)'$, for all $W \in \Ws$.
As $\Pid$ acts transitively upon $\Ws$, the proof is completed.
\end{proof}


\begin{thebibliography}{Bor}
\footnotesize

\bibitem{Ar1}
H. Araki, On the algebra of all local observables, {\sl Prog. Theor.
Phys., \bf 32}, 844--854 (1964).

\bibitem{Ar2}
H. Araki, Some properties of modular conjugation operator
of von Neumann algebras and a non-commutative Radon--Nikodym theorem
with a chain rule, {\sl Pac. J. Math., \bf 50}, 309--354 (1974).

\bibitem{BiWi}
J.J. Bisognano and E.H. Wichmann, On the duality condition for a
Hermitian scalar field, {\sl J. Math. Phys., \bf 16}, 985--1007 (1975).

\bibitem{BiWi2}
J.J. Bisognano and E.H. Wichmann, On the duality condition for
quantum fields, {\sl J. Math. Phys., \bf 17}, 303--321 (1976).

\bibitem{Bor1}
H.-J. Borchers, On structure of the algebra of field operators,
{\sl Nuovo Cim., \bf 24}, 214--236 (1962).

\bibitem{Bor2}
H.-J. Borchers, On the structure of the algebra of field operators, II,
{\sl Commun. Math. Phys., \bf 1}, 49--56 (1965).

\bibitem{BoYn}
H.-J. Borchers and J. Yngvason, On the algebra of field operators:
the weak commutant and integral decompositions of states, {\sl Commun.
Math. Phys., \bf 42}, 231--252 (1975).

\bibitem{BrRo}
O. Bratteli and D.W. Robinson, {\it Operator Algebras and
Quantum Statistical Mechanics I}, Berlin, Heidelberg, New York:
Springer Verlag, 1979. 

\bibitem{BFK}
R. Brunetti, K. Fredenhagen and M. K\"ohler, The microlocal spectrum
condition and Wick polynomials of free fields on curved spacetimes,
{\sl Commun. Math. Phys. \bf 180}, 633--652 (1996).

\bibitem{BS1}
D. Buchholz and S.J. Summers, An algebraic characterization of 
vacuum states in Minkowski space, {\sl Commun. Math. Phys. \bf 155}, 449--458 
(1993).

\bibitem{BFS}
D. Buchholz, M. Florig and S.J. Summers, An algebraic 
characterization of vacuum states in Minkowski space, II: Continuity aspects, 
{\sl Lett. Math. Phys., \bf 49}, 337--350 (1999).

\bibitem{BFS1}
D. Buchholz, M. Florig and S.J. Summers, The second law of
thermodynamics, TCP and Einstein causality in anti-de Sitter space--time,
{\sl Class. Quantum Grav., \bf 17}, L31--L37 (2000). 

\bibitem{BDFS}
D. Buchholz, O. Dreyer, M. Florig and S.J. Summers, Geometric 
modular action and spacetime symmetry groups, {\sl Rev. Math. Phys., \bf 12}, 
475--560 (2000).

\bibitem{BMS1}
D. Buchholz, J. Mund and S.J. Summers, Transplantation of local nets
and geometric modular action on Robertson--Walker space--times, in:
{\it Mathematical Physics in Mathematics and Physics (Siena)} 
(R.~Longo, ed.), {\sl Fields Institute Communications, \bf 30}, 65--81
(2001).

\bibitem{BMS2}
D. Buchholz, J. Mund and S.J. Summers, Covariant and quasi-covariant
quantum dynamics in Robertson--Walker space--times, {\sl Class. Quantum 
Grav., \bf 19}, 6417--6434 (2002).

\bibitem{BS2}
D. Buchholz and S.J. Summers, An algebraic 
characterization of vacuum states in Minkowski space, III: Reflection
maps, {\sl Commun. Math. Phys., \bf 246}, 625--641 (2004).

\bibitem{AdS}
D. Buchholz and S.J. Summers, Stable quantum systems in anti-de Sitter
space: Causality, independence and spectral properties, to appear in:
{\sl J. Math. Phys.}

\bibitem{Dix}
J. Dixmier, {\it Les alg\`ebres d'op\'erateurs dans l'espace Hilbertien},
Paris: Gauthier--Villars, 1969.

\bibitem{DS}
W. Driessler and S.J. Summers, Central decomposition of 
Poincar\'e-invariant nets of local field algebras and absence of spontaneous 
breaking of the Lorentz group, {\sl Ann. Inst. Henri Poincar\'e, \bf 43}, 
147--166 (1985).

\bibitem{DS2}
W. Driessler and S.J. Summers, On the decomposition of relativistic
quantum field theories into pure phases, {\sl Helv. Phys. Acta, \bf 59},
331--348 (1986).

\bibitem{FH}
K. Fredenhagen and R. Haag, Generally covariant quantum field theory
and scaling limits, {\sl Commun. Math. Phys., \bf 108}, 91--115 (1987).

\bibitem{HaKa}
R. Haag and D. Kastler, An algebraic approach to quantum field
theory, {\sl J. Math. Phys., \bf 5}, 848--861 (1964).

\bibitem{Haag}
R. Haag, {\it Local Quantum Physics}, Berlin: Springer-Verlag, 
1992. 

\bibitem{Ju}
J.-P. Jurzak, Decomposable operators application to K.M.S. weights in 
a decomposable von Neumann algebra, {\sl Rep. Math. Phys., \bf 8},
203--228 (1975).

\bibitem{KaRi}
R.V. Kadison and J.R. Ringrose, {\it Fundamentals of the 
Theory of Operator Algebras}, Volume II, Orlando: Academic Press, 1986.

\bibitem{KaWa}
B.S. Kay and R.M. Wald, Theorems on the uniqueness and thermal
properties of stationary, nonsingular, quasifree states on space-times with
a bifurcate Killing horizon, {\sl Phys. Rep., \bf 207}, 49-136 (1991).

\bibitem{La}
L.J. Landau, Asymptotic locality and the structure of local internal
symmetries, {\sl Commun. Math. Phys., \bf 17}, 156--176 (1970).

\bibitem{LaWi}
L.J. Landau and E.H. Wichmann, On the translation invariance of local
internal symmetries, {\sl J. Math. Phys., \bf 11}, 306--311 (1970).

\bibitem{Ma1}
K. Maurin, Mathematical structure of Wightman formulation of quantum
field theory, {\sl Bull. Acad. Polon. Sci. S\'er. sci. math., astr.,
et phys., \bf 11}, 115--119 (1963).

\bibitem{Ma2}
K. Maurin, On some theorems of H.-J. Borchers, {\sl Bull. Acad. Polon. 
Sci. S\'er. sci. math., astr., et phys., \bf 11}, 121--123 (1963).

\bibitem{PuWo}W. Pusz and S.L. Woronowicz, Passive states and KMS states
for general quantum systems, {\sl Commun. Math. Phys., \bf 58}, 273--290 
(1978).

\bibitem{Ra}
M.-J. Radzikowski, {\it The Hadamard Condition and Kay's Conjecture
in (Axiomatic) Quantum Field Theory on Curved Space-Times}, Ph.D. Dissertation,
Princeton University, 1992. 

\bibitem{Ra2}
M.J. Radzikowski, Micro-local approach to the Hadamard condition in
quantum field theory on curved space-time, {\sl Commun. Math. Phys., \bf 179},
529--553 (1996).

\bibitem{ReSch2}
H. Reeh and S. Schlieder, Bemerkungen zur Unit\"ar\"aquivalenz von
lorentzinvarianten Feldern, {\sl Nuovo Cim., \bf 22}, 1051--1068 (1961).

\bibitem{ReSch}
H. Reeh and S. Schlieder, \"Uber den Zerfall der Feldoperatoren im Falle
einer Vakuumentartung, {\sl Nuovo Cim. \bf 26}, 32--41 (1962).

\bibitem{StWi}
R.F. Streater and A.S. Wightman, {\it PCT, Spin and Statistics, and
All That}, Reading, Mass.: Benjamin/Cummings Publ. Co., 1964.

\bibitem{SWi}
S.J. Summers and E.H. Wichmann, Concerning the condition of additivity
in quantum field theory, {\sl Ann. Inst. Henri Poincar\'e, \bf 47},
113--124 (1987).  

\end{thebibliography}
\end{document}